# Pericenter passage of the gas cloud G2 in the Galactic Center

S Gillessen[1], R Genzel[1,2], T K Fritz[1], F Eisenhauer[1], O Pfuhl[1], T Ott[1], M Schartmann[4,1], A Ballone[4,1] and A Burkert[4,1]

[1]Max Planck Institute for Extraterrestrial Physics, PO Box 1312, Giessenbachstr., 85741 Garching, Germany
[2]Departments of Physics and Astronomy, Le Conte Hall, University of California, 94720 Berkeley, USA
[4]Universitätssternwarte der Ludwig-Maximilians-Universität, Scheinerstr. 1, D-81679 München, Germany

We have further followed the evolution of the orbital and physical properties of G2, the object currently falling toward the massive black hole in the Galactic Center on a near-radial orbit. New, very sensitive data were taken in April 2013 with NACO and SINFONI at the ESO VLT[1]. The 'head' of G2 continues to be stretched ever further along the orbit in position-velocity space. A fraction of its emission appears to be already emerging on the blue-shifted side of the orbit, past pericenter approach. Ionized gas in the head is now stretched over more than 15,000 Schwarzschild radii $R_S$ around the pericenter of the orbit, at $\approx$ 2000 $R_S \approx$ 20 light hours from the black hole. The pericenter passage of G2 will be a process stretching over a period of at least one year. The Brackett-$\gamma$ luminosity of the head has been constant over the past 9 years, to within ± 25%, as have the line ratios Brackett-$\gamma$ / Paschen-$\alpha$ and Brackett-$\gamma$ / Helium-I. We do not see any significant evidence for deviations of G2's dynamical evolution, due to hydrodynamical interactions with the hot gas around the black hole, from a ballistic orbit of an initially compact cloud with moderate velocity dispersion. The constant luminosity and the increasingly stretched appearance of the head of G2 in the position-velocity plane, without a central peak, is not consistent with several proposed models with continuous gas release from an initially bound zone around a faint star on the same orbit as G2.

## 1. Introduction

In 2011, we have discovered from our VLT-based, near-infrared observations of the Galactic Center (GC) the small gas cloud G2 falling almost directly onto Sgr A* (Gillessen et al. 2012). It is most prominent in the hydrogen recombination lines Brackett-$\gamma$ and Paschen-$\alpha$, where its emission is unconfused due to the cloud's high Doppler shift. We have also observed the thermal emission of $\approx$ 600 K dust embedded in G2 in L'-band ($\lambda$ = 3.8 μm). The combined data yield a highly eccentric orbit with a pericenter passage in 2013. Between 2004 and 2011 the cloud has developed an increasing velocity gradient, consistent with tidal shearing from Sgr A* along the orbit. We have consolidated this basic picture by additional deep spectroscopy in 2012 (Gillessen et al. 2013). Recently, Phifer et al. (2013) have reported Keck measurements of G2, confirming that the object is moving on a highly eccentric orbit and that it shows a tidal shear. They have shown that using astrometry of the Brackett-$\gamma$ emission instead of L'-band leads to a slightly different orbit with an even higher eccentricity and a pericenter date early 2014. The high eccentricity of e $\approx$ 0.97 is surprising for a gas cloud, although there is an observational bias, because we would have not been able to discern G2 from the ambient gas, if it were not moving with a large line-of-sight velocity.

The nature of the object is enigmatic. The observed phenomenology is that of a non-self-gravitating gas cloud. Attempts to detect a star inside have yielded only upper limits (Gillessen et al. 2012: $m_K$ > 17.8, Phifer et al. 2013: $m_K$ > 20). The compactness of the object in the early 2000's is puzzling (Burkert et al. 2012), since tidal disruption should have long occurred during the orbit, if the cloud had been injected at the apocenter distance of ≥ 1". There are then two options. One is that the cloud actually formed recently, perhaps as a thermal instability in colliding winds of nearby Wolf-Rayet stars (Cuadra et al. 2006), or in the interaction of circum-nuclear gas with the jet from Sgr A* (Gillessen et al. 2013). Alternatively, and perhaps a priori more likely is that the gas reservoir was initially bound to a faint stellar core and then increasing amounts of gas are released and tidally disrupted as the star approaches Sgr A* (Murray-Clay & Loeb 2012, Meyer & Meyer-Hofmeister 2012, Scoville & Burkert 2013, Ballone et al. 2013). Another exciting aspect of G2 is that it potentially will interact with the ambient gas. G2 is moving through the accretion flow of Sgr A* (Yuan et al. 2003, Xu et al. 2006), the density profile of which is increasing inwards like $\approx r^{-1}$. The interaction between G2 and this ambient gas should lead at some point to marked deviations from a purely Keplerian orbit (Schartmann et al. 2012, Anninos et al. 2012), and might also lead to emission in the X-ray (Gillessen et al. 2012) and radio bands (Narayan, Özel & Sironi 2012, Sadowski et al. 2013). The nominal point of closest approach to Sgr A* is at $\approx 2000$ Schwarzschild radii ($R_S$), and hence one might expect that G2 can increase the accretion rate onto Sgr A*. Given the unique opportunity to observe the fate of a gas cloud as it approaches a massive black hole, numerous observing campaigns have been set up to monitor the Sgr A* region in 2013 intensively[2]. Here, we report on our early 2013 re-observations of G2, both in imaging and in unprecedentedly deep integral field spectroscopy.

---

[1] Based on observations collected at the European Southern Observatory, Paranal, Chile; programs 091.B-0081(A), 091.B-0088(A).

## 2. Observations & data reduction

In addition to our data set presented in Gillessen et al. (2012, 2013), we obtained an additional epoch of L'-band imaging, and one epoch of very deep integral field spectroscopy. In the night April 4/5, 2013, we observed the Galactic Center with NACO (Rousset et al. 1998, Lenzen et al. 1998) in L'-band with an image scale of 27 mas pix$^{-1}$. The data are diffraction-limited with a Strehl ratio above 50%. We reduced the data following the standard recipes for L'-band data, constructing the background from the median of the individual frames taken shortly after each other, and which are randomly dithered with respect to each other. We used the 'starfinder' tool (Diolaiti et al. 2000) to extract the point spread function centered on Sgr A* from the combined image, followed by a moderately deep Lucy-Richardson deconvolution (Lucy 1974).

Between April 5, 2013 and April 18, 2013, we observed the central arcsecond with the adaptive-optics based integral field spectrometer SINFONI (Eisenhauer et al. 2003, Bonnet et al. 2004) for a total of 1480 minutes (24:40 hours) on-source integration time. We used the combined H+K-band setting with a spectral resolution of $\approx 1500$, and a pixel scale of $12.5 \times 25$ mas pix$^{-1}$. We used the closest optical guide star 22" northeast of Sgr A* with mag$_R$ = 14.65, which makes the data quality depend strongly on the atmospheric conditions. After the usual reduction steps (detector calibration, cube reconstruction, wavelength calibration) we used the FWHM of the star S2 at 2.2 µm as a quality criterion for selecting cubes. We demanded that the FWHM < 6.5 pix (81 mas), which yielded a total of 1160 minutes (19:20 hours) on-source data. This is the deepest integral-field exposure of Sgr A* and its immediate vicinity ever. The correction for Earth's atmosphere was done using the spectrum of S2, in which we have manually fitted and removed the few hydrogen and helium lines.
Overall, we detect G2 also in 2013 at high significance, although its tidal stretch has developed further, lowering the surface brightness significantly.

## 3. Analysis

We do not detect G2 in our new deconvolved L'-band image (figure 1). Using 'starfinder' we detect a point-like source, but only roughly at the expected position, The deconvolved image shows that the confusion at the expected position is too high to rely on the astrometry. Hence, we don't get a new L'-band astrometric data point, leaving us with the previous set of 14 L'-band based positions.

Beyond the new SINFONI data set, we were also able to find back G2 in more archival SINFONI data sets. We were able to derive additional radial velocities in 2006 and 2007, together with the new 2013 epoch and an observation from September 2012, this yields a total of 15 radial velocity data points. To this end, we extracted the spectra from the cubes by selecting the on-region as the area in which G2's line emission is well detectable above the background, and the off-region as ring around the on-region. The uncertainties in the resulting velocities are due to the freedom in selecting the on- and off-regions, and due to the fitting errors.

---

[2] See: https://wiki.mpe.mpg.de/gascloud

Following the analysis of Phifer et al. (2013), we derived Brackett-γ astrometry from our SINFONI cubes. G2's position was measured in channel maps centered on the respective measured radial velocity. In each cube, we also measured the position of all S-stars at the same wavelength where G2 is visible. We know the astrometric positions of these stars with high precision (Gillessen et al. 2009), from which we set up a local transformation of pixel coordinates to astrometric positions. Eleven of our cubes contain a sufficient number of stars to derive a position for G2. The associated errors follow from both the positional error from measuring the position of G2 in the cube, and the transformation error, which we estimate by comparing the transformed positions of the stars with their nominal positions. The one-dimensional errors vary between 1.5 mas and 8 mas. Note that due to the differences in defining the coordinate system, it is not possible to directly compare the positions in Phifer et al. (2013) with our Brackett-γ astrometry. Also, one should note that SINFONI uses an image slicer to split the field of view optically in front of the spectrograph. In contrast, the data from Phifer et al. (2013) were obtained with OSIRIS, which uses a lenslet array. SINFONI is thus more prone to suffer from optical distortions than OSIRIS, although both instruments have not been designed for astrometry.

In summary, we have two different data files for G2: Both share the same radial velocity data, but one contains the L'-band based astrometry, the other the Brackett-γ derived positions.

We use the same technique as in Gillessen et al. (2013) to extract position-velocity-diagrams from the data cube: Namely we project the orbit into the cube and extract the spectra along this virtual, curved slit. Given our two data files, we get two different orbits, and hence we try both such curved slits. It turns out that the actual pixel sets selected for the two orbits are almost identical. In the time range from 2002 to 2013 over which we observed G2, the projections differ by at most one pixel, and hence the choice does not matter. Also the tail and bridge structures are comparable for the two projections, such that the only difference is a non-linear transformation between the two ways of defining the positional axis along the orbital trace. We use (integer) slit widths between 5 and 9 pixels, and select after extraction the diagram with the best signal-to-noise ratio, which yields a slit width of 6 pixels. As in the previous years, we detect G2 not only in Brackett-γ, but also in Helium-I ($\lambda = 2.058$ μm) and Paschen-α ($\lambda = 1.875$ μm, which is between the H- and the K-band). Hence, we can co-add the position-velocity-diagrams of the three lines after re-binning to the same wavelength scale, using the noise in an empty region of each diagram to determine the respective weights. For the blue-shifted part of the position-velocity-diagram, we cannot use the Paschen-α line due to atmospheric absorption residuals. The signal-to-noise ratio is thus lower in that part. Finally, we smooth the position-velocity-diagram with a Gaussian kernel with FWHM = 2 pix (figure 2).

In addition, we have been able to get position-velocity-diagrams from our archival SINFONI data also for 2006 and 2010. The data sets are not as deep as our 2012 and 2013 exposures, yet they complement the set of position-velocity-diagrams. In total, we can present a time series of seven such diagrams, nicely illustrating the evolution of G2 from 2004 to 2013 (figure 3).

3. Results

The position-velocity-diagram shows a very elongated structure. The fastest components appear at a redshift of ≈ 3000 km s$^{-1}$. The brightest part of the head of G2 is at 2180 ± 50 km s$^{-1}$. The tail structure at +700 km s$^{-1}$ is also detected again, as well as the bridge connecting head and tail, and thus suggesting a physical connection between head and tail. The best-fitting orbit passes through head, bridge and tail in the position-velocity-diagram. Compared to 2012, the tidal shear of the head has increased further, as visible in the time series of Brackett-γ line profiles in figure 4. Fitting Gaussian functions to the line profiles yields the evolution of the width as shown in figure 5, left panel. Over the last nine years the line has continued to steadily gain in width[3]. Figure 3 shows that this increase is due to the increasing velocity gradient, and not due to an increased internal random motion of the gas.

Probably as a result of the increased tidal shear, the surface brightness of the line emission of G2's head has decreased dramatically. The signal-to-noise ratio of individual, bright pixels in our 20-hour integration from 2013 is smaller than for our 11-hour data set from 2012. The total Brackett-$\gamma$ luminosity of the red-shifted head remained constant, however, at around $2 \times 10^{-3}$ $L_\odot$, with an error of at least 20% (figure 5, middle panel). Hence, we do not detect any signs of heating of the gas, which would lower the total Brackett-$\gamma$ luminosity. Also the line ratios Paschen-$\alpha$/Brackett-$\gamma$ and Helium-I/Brackett-$\gamma$ have not changed since 2008, the first year for which we can reliably measure them (figure 5, right panel).

Most exciting is the detection of gas emission at a blue shift of $\approx -3000$ km s$^{-1}$ along the orbit at a position after pericenter (figure 2). This emission is clearly detected in Brackett-$\gamma$, and tentatively also in Helium-I. We cannot see it in Paschen-$\alpha$, since that part of the spectrum is hidden behind atmospheric features. The integrated luminosity of that part is a few ten percent of the head structure. We conclude that the first parts of G2 have already passed pericenter. Given the extension of the head and the presence of post-pericenter gas, it is not adequate to think of a single pericenter date. The pericenter passage of G2 is an extended event, the intrinsic duration of which we estimated in Gillessen et al. (2013) from a test particle model to be roughly one year.

G2 appears spatially extended (figure 6). Its intrinsic effective radius is smaller than the diffraction limit (as determined from the S2 continuum in the respective same data sets), and hence a size measurement is very uncertain due to the subtraction in squares. In the data sets with high signal-to-noise ratio (2008, 2010, 2011, 2012, 2013) we can measure a finite size of G2. We obtain an intrinsic Gaussian FWHM size of 42 ± 10 mas, compared to our size estimate of 30 mas in Gillessen et al. (2012). The error is dominated by the epoch-to-epoch variation, however we find no evidence for a consistent trend of size with time.

Phifer et al. 2013 have noted that there are systematic differences between the L'-band and Brackett-$\gamma$ based orbits. We confirm this in our data set. Also our Brackett-$\gamma$ orbit has a later pericenter date (2014.25) and a larger eccentricity (0.9762) than the one which we obtain when using the L'-band data instead. However, the eccentricity of our Brackett-$\gamma$ based orbit is not as extreme as in Phifer et al. (2013), and correspondingly the pericenter distance with 2400 $R_S$ remains closer to our earlier estimates. Another difference to the Keck-based work is that our new Brackett-$\gamma$ based orbit continues to be coplanar with the clockwise stellar disk: It is 20° tilted with respect to the best fitting inner disk of Bartko et al. (2009), and 16° with respect to the disk orientation of Lu et al. (2009). The angular momentum vectors for the eight disk stars, for which we have explicit orbital solution deviate on average by 12° ± 6° from the disk direction given in Bartko et al. (2009). These authors estimated a Gaussian FWHM of 20° for the disk thickness.

It is interesting to compare the different orbits in how well they describe the measured radial velocities (figure 7). The orbit from Phifer et al. (2013) does not match well our new radial velocity, since it would have predicted a value of 2500 km s$^{-1}$ by now. Our data set yields both an L'-band and a Brackett-$\gamma$ astrometry based orbit, both of which can be fit to match our 2013 radial velocity. The L'-band based orbit puts our current measurement close to the tip of the radial velocity curve, just before it swings to very negative values. The strength of this radial velocity model is that it captures the apparent change of curvature in the radial velocity curve. However, as pointed out by Phifer et al. (2013), the L'-band positions might be biased. Also the Brackett-$\gamma$ based astrometry yields a viable radial velocity model. It evolves less sharply than the orbit from Phifer et al. (2013). However, it does not describe well the radial velocities measured in 2012, where it somewhat underpredicts the actual redshift. Table 1 compares the different orbit estimates.

---

[3] The values given in Gillessen et al. (2013) are larger than the numbers here due to a different selection of the extraction region, which in the presence of the velocity shear systematically changes the measured width. Both sets are internally consistent.

## 4. Discussion

Figure 2 shows that the blue-shifted tip of the radial velocity curve for the Brackett-γ based orbit matches the morphology of the blue-shifted gas in the position-velocity-diagram reasonably well, while the L'-band based orbit is a poorer description. This statement also holds, when considering the test particle simulations for the two orbits (figure 8). To that end we placed a cloud of test particles on the respective orbit, starting at t = 2000.0 with a Gaussian FWHM of the velocity dispersion of 120 km s$^{-1}$ and a FWHM of the spatial extent of 42 mas. An additional radial velocity measurement in 2013 will narrow down the possible parameter space of orbits considerably, as is obvious from figure 7. However, it is not clear that a single orbit continues to be an adequate description of the gas dynamics of G2. Detecting gas post-pericenter could also mean that the brightest part of the red-shifted head is not well comparable anymore to the head in 2012. G2 might have entered the phase during which it matters, which part of the original cloud one tries to trace. Further attempts to determine the orbit better would then be prone to fail.

A hint that this is actually the case is the 2013 measured radial velocity that appears to have increased less since 2012 than what one might have expected for the Brackett-γ based orbit. Calculating the difference between mean radial velocity of the redshifted test particles used in figure 8 and the orbit yields an offset of 200 ± 20 km s$^{-1}$ for both orbit estimates. This is because the fastest particles have already moved to the blue-shifted side and miss in the mean. Hence, the test particle model suggests that the 2013 measurement is biased to too small values, and conversely, the measured value suggests that some gas has moved to the blue-shifted side already.

Figure 9 shows the fraction of gas that has not yet passed pericenter as a function of time for the different orbits. Both the Brackett-γ based orbit and the one from Phifer et al. (2013) predict that only around 7% of the gas has already swung by Sgr A* in April 2013. The L'-band orbit, on the other hand, predicts that 25% of the gas already have passed pericenter. The latter would agree better with the detection of a significant flux component at the blue-shifted part of the position-velocity diagram in figure 2. However, the dynamic signature does not match as well as for the Brackett-γ based orbit.

Overall we think that our Brackett-γ based orbit is the best estimate. As pointed out by Phifer et al. (2013) it is less prone to astrometric biases than the L'-band based one, and our SINFONI data seem to suggest that indeed this orbit comes closer to reality. Then the somewhat high luminosity of the blue-shifted gas needs to be explained by a combination of a larger size of G2 along the orbit, a possibly higher emissivity of the gas post-pericenter as could result from a shock-induced, increased density, and eventually hydrodynamics affecting the orbital motion. A more stretched out spatial figure of G2 is not surprising, if it formed before our arbitrary starting date (Burkert et al. 2012).

Given the large tidal shear, we can test a prediction for models of G2 that contain a central source of gas, and for which the outflowing gas is photo-ionized by the UV radiation field in the central arcsecond, such as proposed by Murray-Clay & Loeb (2012). With a central source, the radial profile of the gas density around the star should behave roughly like r$^{-2}$. We have simulated in figure 10 how such a source would look like in the position-velocity-diagram, where we also compare with the test particle model described above. We implemented the r$^{-2}$-profile simply by using the same type of Gaussian simulation, but weighting each test particle at the time steps of interest by r$^{-2}$. We have quadrupled the initial spatial FWHM in order to match the measured width of the data in the earlier years. The simulation shows that the velocity gradient would be considerably steeper than what we observed. Comparing with figure 3, we conclude that the data favor the pure test particle cloud model.

The model of Scoville & Burkert (2013) predicts that the emission predominantly arises from the cold bow shock which for typical wind velocities should be at 10 – 100 AU from the central source, corresponding to an observable intrinsic radius of 1.2 – 12 mas, which is at least a factor 2 smaller than our size estimate. In contrast, Ballone et al. (2013) derive the parameters of their standard model by directly comparing their simulated position-velocity-diagrams as well as the Brackett-γ luminosity evolution with the observations. Therefore, the size evolution of that model is in agreement with the observations by construction.

The test particle simulation, with the updated orbit and spatial FWHM size, continues to be a surprisingly good description of the dynamics of G2. We have also tested that the initial velocity dispersion is required. Repeating the simulation with zero dispersion yields too small velocity widths in the first years. While in 2004 and 2006 the measured values are close to the instrumental resolution, the 2008 measured value with 240 km s$^{-1}$ is well above the FWHM resolution of $\approx$ 120 km s$^{-1}$. Here, the zero dispersion simulation would predict an intrinsic line width of 34 km s$^{-1}$ only, which would still be unresolved. Physical models that could agree with the test particle model include the debris from stellar wind collision, or a cooling instability in the accretion flow (Burkert et al. 2012). In the former picture, the high eccentricity of the orbit is puzzling, while in the latter it is natural. The compactness of G2 is surprising in either case. On the other hand, the orientation of the orbit is more natural if the origin of G2 is connected to the young, massive stars in the clockwise disk.

Our data do not show any clear evidence of hydrodynamic effects, although a more detailed comparison with simulations might reveal that still. Also, the simulations show that there are no significant deviations from the test particle model to be expected before pericenter passage. Only after that, the effects should become visible. Due to the decreasing timescale for Kelvin-Helmholtz instabilities, the gas starts to mix efficiently and gets much more spread out then. In particular the position-velocity-diagram should show a structure that is oriented perpendicular to the orbit, and which extends into a stream feeding the innermost accretion zone of Sgr A* (figure 9 of Schartmann et al. 2012). The large extension and correspondingly low surface brightness of the gas, however, will make such a detection difficult.

**References**

Anninos, P., Fragile, P. C., Wilson, J., & Murray, S. D. 2012, ApJ, 759, 132
Ballone, A., et al. 2013, arXiv:1305.7238
Bartko, H., et al. 2009, ApJ, 697, 1741
Bonnet, H., et al. 2004, ESO Messenger, 117, 17
Burkert, A., Schartmann, M., Alig, C., Gillessen, S., Genzel, R., Fritz, T. K., & Eisenhauer, F. 2012, ApJ, 750, 58
Cuadra, J., Nayakshin, S., Springel, V., Di Matteo, T. 2006, MNRAS, 366, 358
Diolaiti, E., Bendinelli, O., Bonaccini, D., Close, L., Currie, D., & Parmeggiani, G. 2000, A&AS, 147, 335
Eisenhauer, F., et al. 2003, Proc. SPIE, 4841, 1548
Gillessen, S., Eisenhauer, F., Trippe, S., Alexander, T., Genzel, R., Martins, F., & Ott, T. 2009, ApJ, 692, 1075
Gillessen, S., Genzel, R., Fritz, T. K., Quataert, E., Alig, C., Burkert, A., Cuadra, J., Eisenhauer, F., Pfuhl, O., Dodds-Eden, K., Gammie, C. F., Ott, T. 2012, Nature, 481, 51
Gillessen, S., Genzel, R., Fritz, T. K., Eisenhauer, F., Pfuhl, O., Ott, T., Cuadra, J., Schartmann, M., Burkert, A. 2013, ApJ, 763, 78
Lenzen, R., Hofmann, R., Bizenberger, P., & Tusche, A. 1998, Proc. SPIE, 3354, 606
Lu, J., Ghez, A. M., Hornstein, S., Morris, M., Becklin, E., & Matthews, K. 2009, ApJ, 690, 1463
Lucy, L. B. 1974, AJ, 79, 745
Meyer, F. & Meyer-Hofmeister, E. 2012, A&A, 546, L2
Murray-Clay, R. A. & Loeb, A. 2012, Nat. Commun. 3:1049 doi: 10.1038/ncomms2044
Narayan, R., Özel, F. & Sironi, L. 2012, ApJ, 757, L20
Phifer, K. et al. 2013, arXiv:1304.5280
Rousset, G., et al. 1998, Proc. SPIE, 3353, 508
Sadowski, A. et al. 2013, MNRAS, 432, 478
Schartmann, M., Burkert, A., Alig, C., Gillessen, S., Genzel, R., Eisenhauer, F., & Fritz, T. K. 2012, ApJ, 755, 155
Scoville, N. & Burkert, A. 2013, ApJ, 768, 108
Xu, Y.-D., Narayan, R., Quataert, E., Yuan, F., Baganoff, F. K. 2006, ApJ, 640, 319

**Figures**

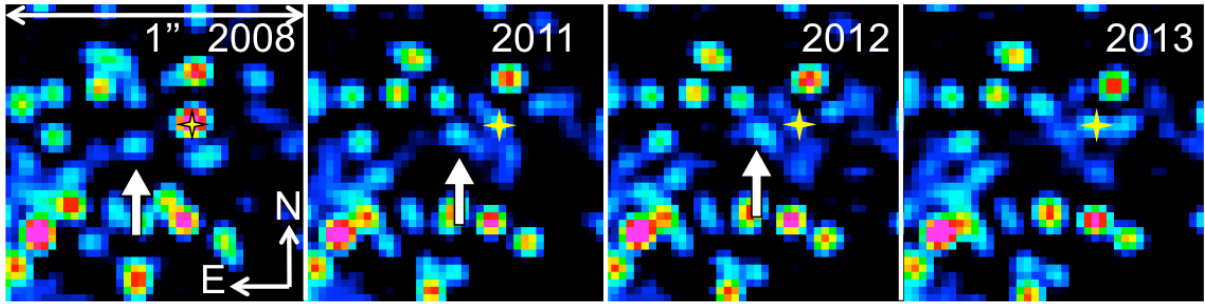

Figure 1: Series of deconvolved L'-band images of the central arcsecond. G2 is unambiguously detected up to 2012, but not in 2013, where the image does not show a source. Starfinder detects a source but only at roughly the expected position. The high confusion around Sgr A* and G2's lowered surface brightness appear to hamper a detection.

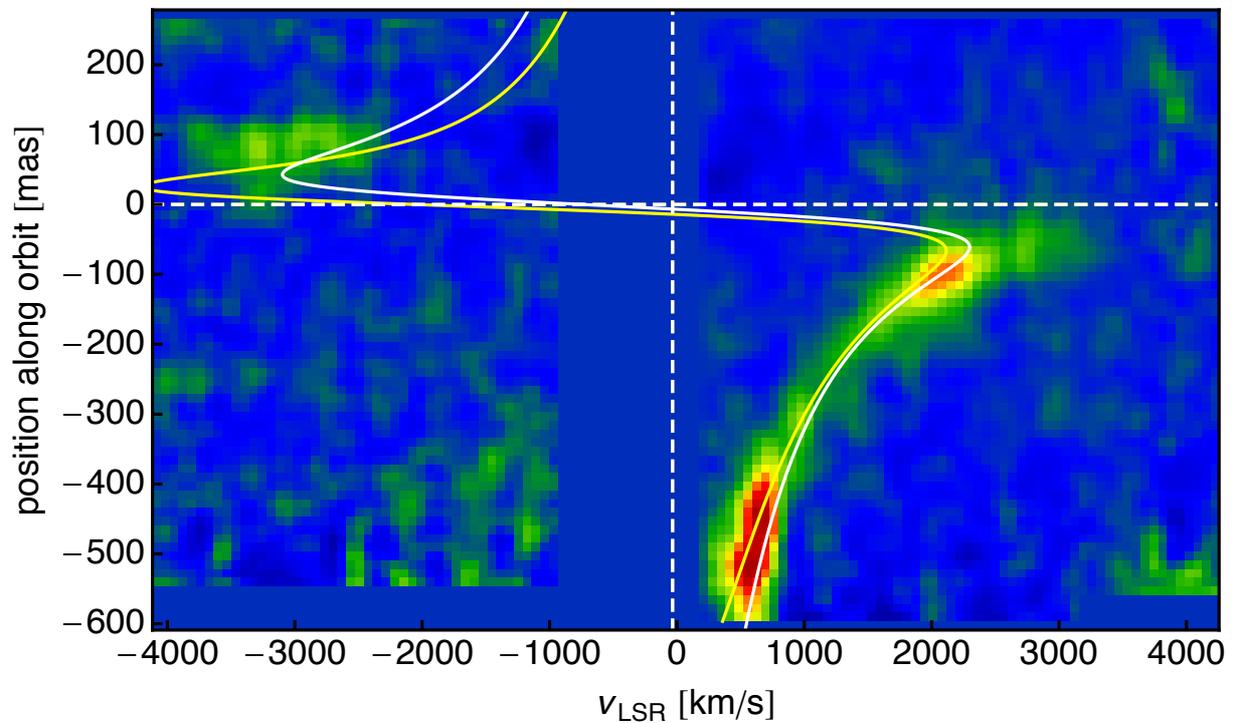

Figure 2: Position-velocity-diagram of G2, extracted from our April 2013 SINFONI data set along the orbit projected into the cube. This diagram is a co-add around the lines Brackett-$\gamma$, Helium-I, and Paschen-$\alpha$. The yellow line delineates the L'-band based orbit, the white line the Brackett-$\gamma$ based one.

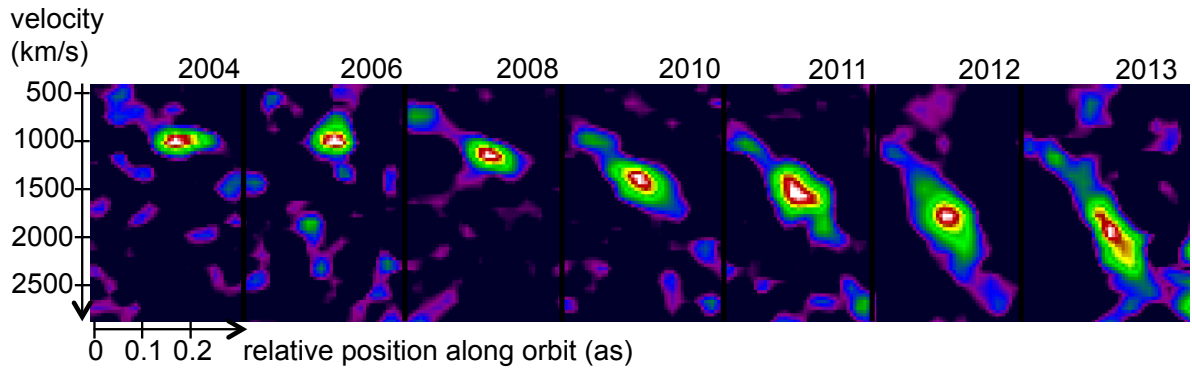

Figure 3: Series of position-velocity-diagrams from 2004 to 2013 extracted from SINFONI data, scaled to identical peak luminosities. For illustration purposes, the original data were resampled in each axis to twice as many pixels as the original, and the diagram was smoothed with a Gaussian kernel of FWHM = 2 pix. The evolution of the tidal shear is obvious.

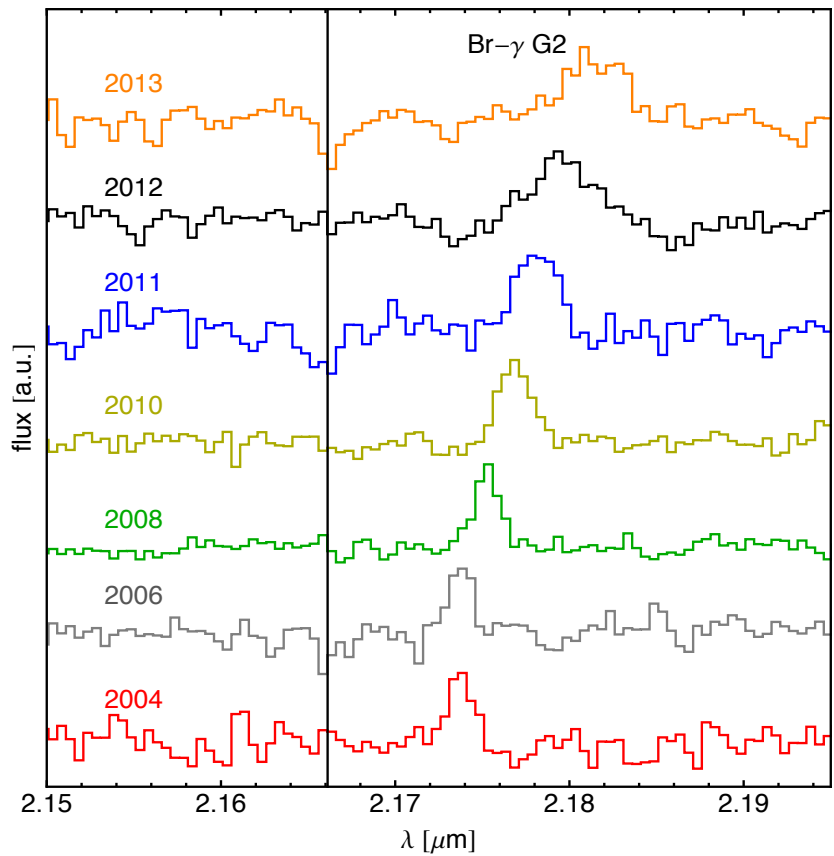

Figure 4: Series of spectra around Brackett-γ of G2 obtained from our SINFONI data set. All spectra are scaled to the same peak flux. The increase in line width with time is obvious. The integrated line flux is constant within errors over time. The vertical line marks the rest wavelength of Brackett-γ.

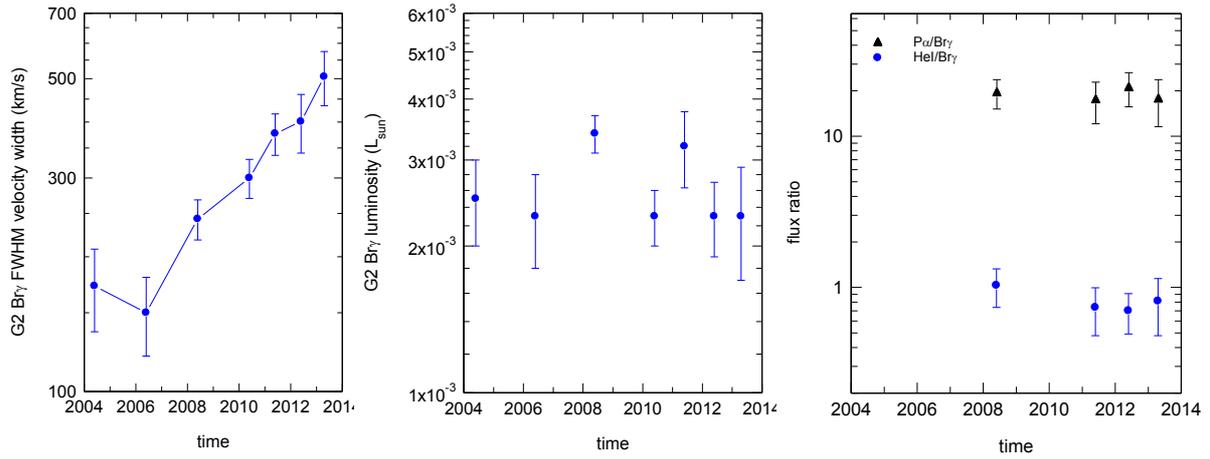

Figure 5: Properties of G2. Left: The temporal evolution of the Brackett-γ FWHM line width shows a continuous increase. The instrumental line width is around 120 km s$^{-1}$, close to the measured values in 2004 and 2006. Middle: The Brackett-γ luminosity is consistent with being constant over time. Right: Also the line ratios Helium-I / Brackett-γ and Paschen-α / Brackett-γ remain constant as a function of time.

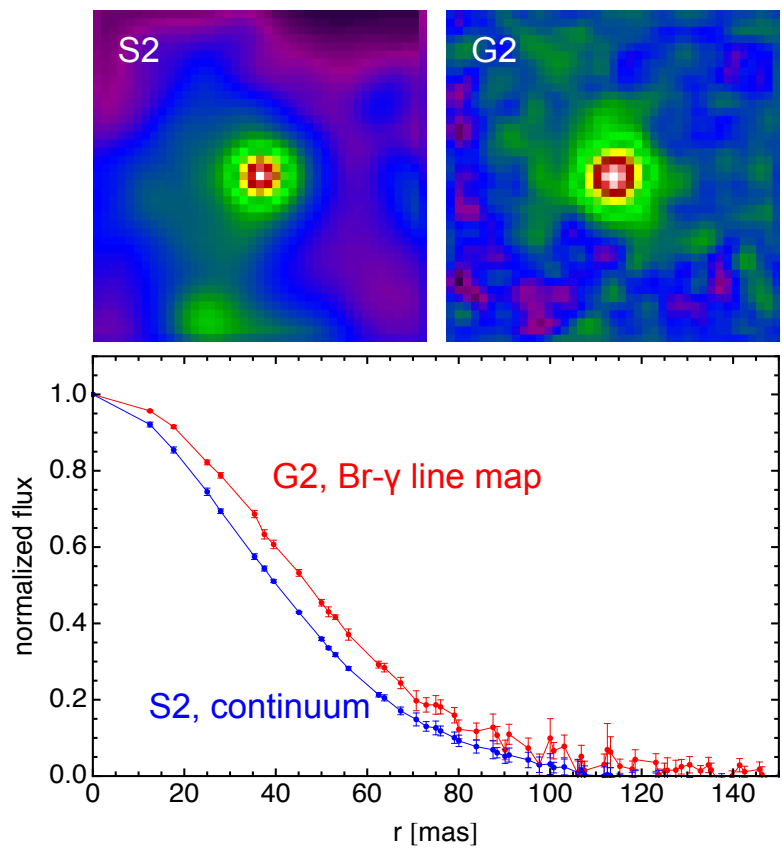

Figure 6: A comparison of the mean radius of G2 with the size of the star S2 shows that G2 is marginally extended. Top left: Continuum emission of S2 at the wavelength of G2's Brackett-γ line as measured with SINFONI, averaged over the years 2008, 2010, 2011, 2012, 2013 and corrected for nearby stars. Top right: The same average of the Brackett-γ line emission of G2. Bottom: Comparison of the two radial flux profiles.

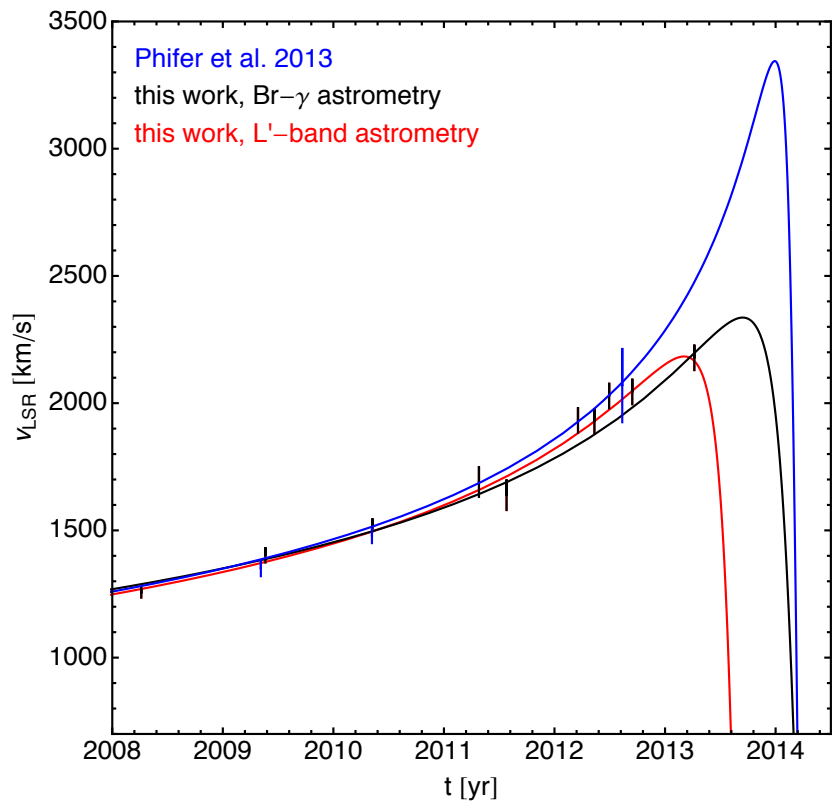

Figure 7: Comparison of the different orbits for G2 with the radial velocity data.

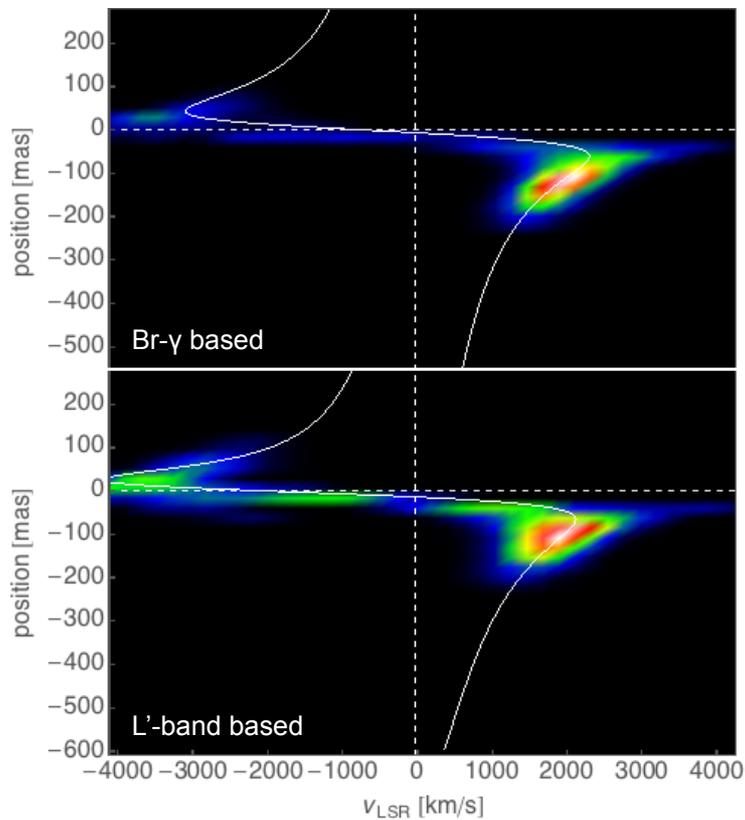

Figure 10: The test particle simulations for the two orbit estimates, evaluated at the April 2013 epoch. Top: The Brackett-γ based orbit. Bottom: The L'-band based orbit. Comparison with figure 3 shows that the Brackett-γ based orbit resembles the observations more closely.

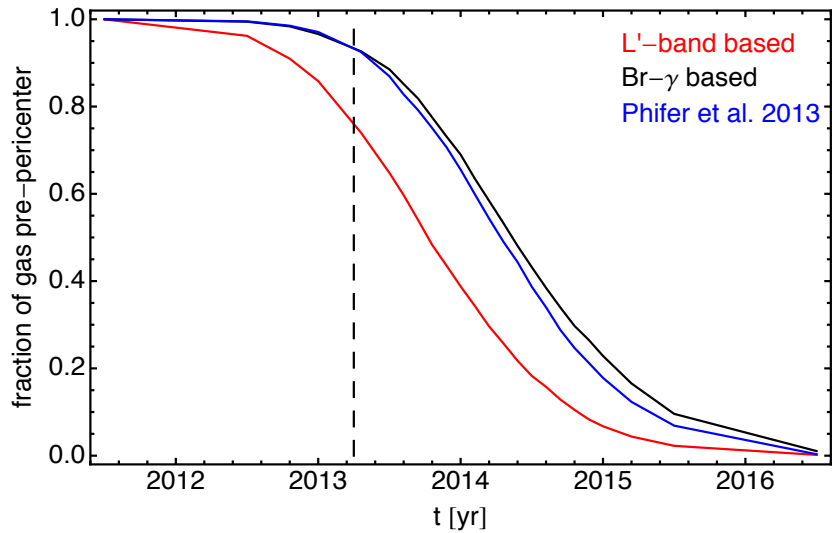

Figure 9: Evolution of the fraction of gas that has not yet passed pericenter for the test particle model. The dashed line marks the epoch of our 2013 observation.

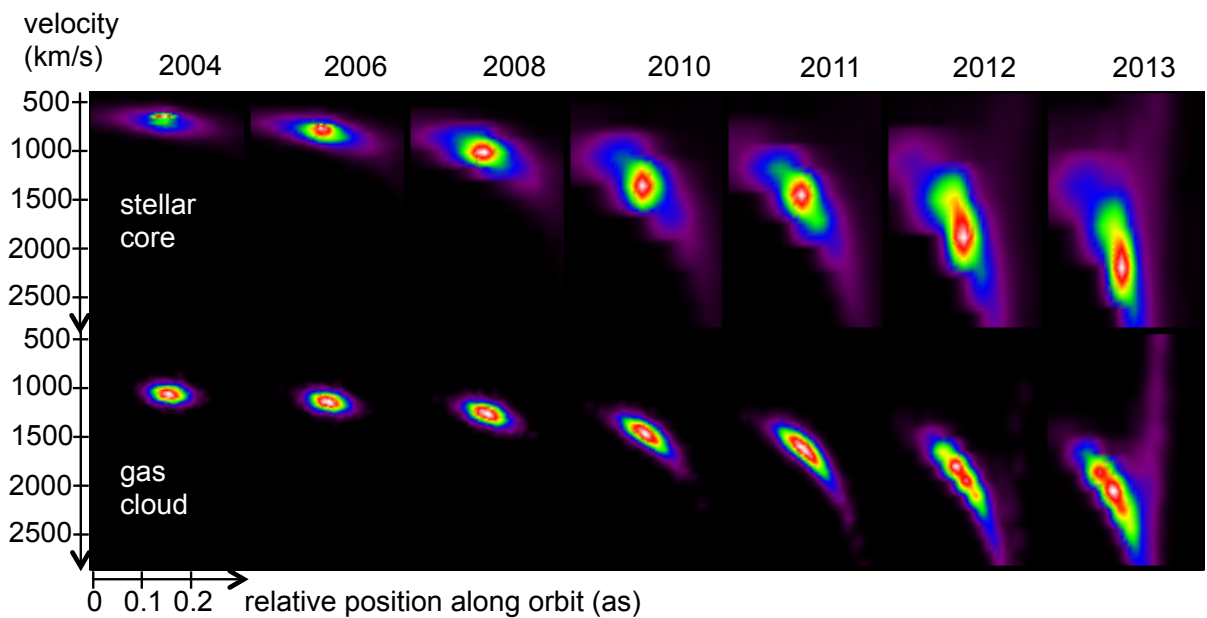

Figure 10: Test particle simulations of the evolution of the position-velocity-diagram. Top: Assuming a radial profile of $r^{-2}$ as might occur in a model with a central gas supply inside of G2 leads to a very steep velocity gradient. Bottom: A Gaussian cloud with a velocity FWHM of 120 km s$^{-1}$ and spatial FWHM 42 mas at t = 2000.0 yields a better description of the observational data (figure 3).

| | Gillessen ea. 2012 | Gillessen ea. 2013 | Phifer ea. 2013 | L'-band based | Br-γ based |
|---|---|---|---|---|---|
| a [mas] | 521 ± 28 | 666 ± 39 | 880 | 684 ± 55 | 1048 ± 247 |
| e | 0.9384 ± 0.0066 | 0.9664 ± 0.0026 | 0.9814 ± 0.0060 | 0.9698 ± 0.0031 | 0.9762 ± 0.0074 |
| i [°] | 106.55 ± 0.88 | 109.48 ± 0.81 | 121 ± 3 | 110.2 ± 1.4 | 118.1 ± 2.0 |
| Ω [°] | 101.5 ± 1.1 | 95.8 ± 1.1 | 56 ± 11 | 94.5 ± 1.8 | 81.9 ± 4.3 |
| ω [°] | 109.59 ± 0.78 | 108.50 ± 0.74 | 88 ± 6 | 108.6 ± 1.2 | 97.2 ± 2.2 |
| $t_0$ [yr] | 2013.51 ± 0.04 | 2013.69 ± 0.04 | 2014.21 ± 0.14 | 2013.72 ± 0.05 | 2014.25 ± 0.06 |
| P [yr] | 137 ± 11 | 198 ± 18 | 276 ± 111 | 206 ± 15 | 391 ± 66 |
| $p_0$ [$R_s$] | 3100 | 2200 | 1500 | 2000 | 2400 |

Table 1: Comparison of orbital elements from the literature and our new estimates. We favor the Brackett-γ based orbit in the right-most column.